# RADIATION PROTECTION ISSUES AFTER 20 YEARS OF LHC OPERATION

D. Forkel-Wirth, M. Magistris, S. Roesler, C. Theis, L. Ulrici, H. Vincke and Hz. Vincke,
CERN DGS-RP, Geneva, Switzerland

*Abstract*

Since November 2009, the LHC commissioning progresses very well, both with proton and lead beams. It will continue in 2011 and nominal LHC operation is expected to be attained in 2013. In parallel, plans for various LHC upgrades are under discussion, suggesting a High-Luminosity (HL) upgrade first and a High-Energy (HE) upgrade in a later state. Whereas the upgrade in luminosity would require the modification of only some few key accelerator components like the inner triplets, the upgrade in beam energy from 7 TeV to 16.5 TeV would require the exchange of all dipoles and of numerous other accelerator components.

The paper gives an overview of the radiation protection issues related to the dismantling of LHC components prior to the installation of the HE-LHC components, i.e. after about 20 years of LHC operation. Two main topics will be discussed: (i) the exposure of workers to ionizing radiation during the dismantling of dipoles, inner triplets or collimators and experiments and (ii) the production, conditioning, interim storage and final disposal of radioactive waste.

## EXPOSURE OF WORKERS TO IONIZING RADIATION

Dismantling of accelerator components from hadron accelerators implies the exposure of workers to ionizing radiation. The ionizing radiation ($\beta,\gamma$) is caused by the radioactive decay of spallation induced radionuclides produced inside the components and their surroundings during beam operation. The level of induced radioactivity is a function of the chemical composition of the component, of the beam particle type and energy, of the beam losses (accelerator) and of luminosity (experiment detectors).

Prior to any dismantling work, a risk analysis has to be performed. Usually, ambient dose equivalent rates and levels of induced activity are measured after the beam stop and fed into the overall job and dose planning for dose optimization. However, the risk analysis for the dismantling of LHC components in 20 years time can only be based on the results of Monte Carlo simulations. Indeed, the comparison with and the extrapolation from measurements are not yet possible, as the activation and the radiation levels in the LHC are still very low.

Most of the FLUKA calculations for the LHC were performed assuming 180 days of operation at nominal beam conditions. The extrapolation up to 20 years of LHC operation requires additional inputs, such as the radiation protection relevant LHC parameters (beam energy, beam intensity, luminosity) for 20 years of LHC operation (see Table 1) determining the build-up of long lived isotopes (e.g. $^{60}$Co, $^{22}$Na), as well as their contribution to the ambient dose equivalent rate. For this purpose a generic study was performed: the activation of a simplified magnet (iron or steel cylinder) was simulated to estimate the contribution of the long-lived radionuclides to the ambient dose equivalent rate assuming 180 days, 5 years and 20 years of LHC operation. The 5 (20) years of LHC beam operation were approximated by assuming 5 (20) times one year of 180-day irradiation and 185-day shut-down. The simulation took into account the chemical composition of the material used for LHC components. As an example, the composition of steel for the LHC dipoles is listed in Table 2.

Figure 1 gives the FLUKA results per proton at 7 TeV, for the three different irradiation times and followed by 4-month cooling time.

Table 1: LHC parameters relevant for the calculation of induced radioactivity at the various stages of LHC operation

| LHC Phase | Energy (TeV) | Beam Intensity (pr. per beam) | Peak Luminosity ($cm^{-2} s^{-1}$) | Year |
|---|---|---|---|---|
| Commission. | 3.5 | $5.1 \cdot 10^{13}$ | $2 \cdot 10^{32}$ | 2010 |
| Commission. | 3.5 | $1.5 \cdot 10^{14}$ | $1 \cdot 10^{33}$ | 2011 |
| Nominal | 7 | $3.2 \cdot 10^{14}$ | $1 \cdot 10^{34}$ | 2013 |
| Ultimate | 7 | $4.7 \cdot 10^{14}$ | $2.3 \cdot 10^{34}$ | 2017 |
| HL-LHC | 7 | $4.7 \cdot 10^{14}$ | $5 \cdot 10^{34}$ | 2021 |
| HE-LHC | 16.5 | $2.5 \cdot 10^{14}$ | $2 \cdot 10^{34}$ | >2030 |

Table 2: Chemical composition of steel used for the LHC dipoles

| Steel composition | | | | | |
|---|---|---|---|---|---|
| Elem. | Wt-% | Elem. | Wt-% | Elem. | Wt-% |
| Fe | 63.09 | S | 0.00 | Mo | 0.09 |
| Cr | 17.79 | Cu | 0.09 | C | 0.10 |
| Ni | 6.50 | O | 0.00 | W | 0.01 |
| Mn | 11.43 | Ti | 0.01 | P | 0.02 |
| Si | 0.38 | V | 0.07 | Nb | 0.01 |
| N | 0.31 | Co | 0.11 | | |

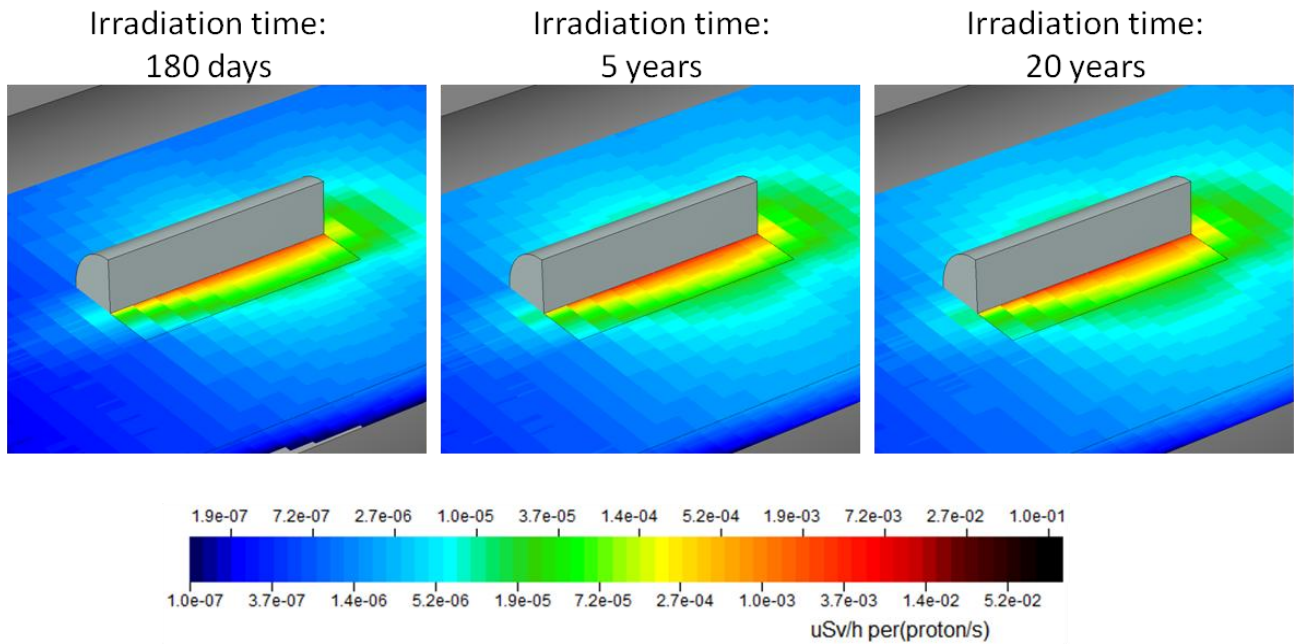

Figure 1: Results of generic calculations of ambient dose equivalent rates (per proton) after irradiation of a simplified magnet (steel cylinder) at 7 TeV and 4-month cooling. Three different irradiation times are considered: 180 days, 5 years and 20 years of LHC operation.

The graphic in Fig. 2 compares the ambient dose equivalent rates found along the magnet for the three scenarios. The results indicate an increase of the ambient dose equivalent rate by a factor of about 1.7 between 1 year and 5 years of operation and about a factor 2 between 1 year of operation and 20 years of operation.

To allow the extrapolation from presently available results of FLUKA calculations, it was assumed that the LHC technical installation will not be modified and the beam loss pattern will not change over the next 20 years. Under these assumptions, the ambient dose equivalent rates depend on beam energy ($E^{0.8}$), luminosity (experiments, inner triplet), beam intensity (arcs, collimators) and total number of protons.

Three examples will be given for extrapolated ambient dose equivalent rates:

- **LHC ARCs**: the ambient dose equivalent rates were calculated for nominal operation, assuming 180 days of operation, a beam gas interaction rate of $2.4 \times 10^4$ protons/m/s (both beams) at 7 TeV and which corresponds to a $H_2$-equivalent beam gas density of $4.5 \times 10^{14}$ m$^{-3}$. Under these assumptions, the ambient dose equivalent rates inside the arc magnets and close to the beam line will reach 20 µSv/h after 1-month cooling, about 300 to 400 nSv/h at the surface of the cryostate and about 200 nSv/h in the aisle. After 20 years of LHC operation, in particular after operation of LHC as HL-LHC, the expected ambient dose equivalent rates are estimated to be about a factor of 3 higher.
- **Inner triplet**: the ambient dose equivalent rate at the surface of the cryostat will be in the order of 600 µSv/h after 5 years of operation under nominal conditions and 4-month cooling. After 10 years of HL-LHC, the ambient dose equivalent rate at the surface of the cryostate will reach about 1 mSv/h after 4-month cooling. Inside the magnets the dose rates will be higher and of a different order of magnitude.
- **Collimator Region**: After one year of operation at the nominal beam intensities the ambient dose equivalent rate in the aisle will reach some 10 to 100 µSv/h, and close to the collimator it will be 100 µSv/h to 1 mSv/h after 4-month cooling. After 20 years of LHC operation and the same cooling time, the dose rates are estimated to be about a factor of 3 higher and reach up to 3 mSv/h close to the collimator.

The removal of dipoles will imply destructive work, like for example cutting the beam pipes and splices. This work entails a risk of contamination. Adequate techniques will have to be developed already for the splice-repair campaign, which is foreseen for 2012. The dose to the workers has also to be optimized for the transport of components: passing the collimators in Point 3 and Point 7 may result in non-negligible doses to the transport team.

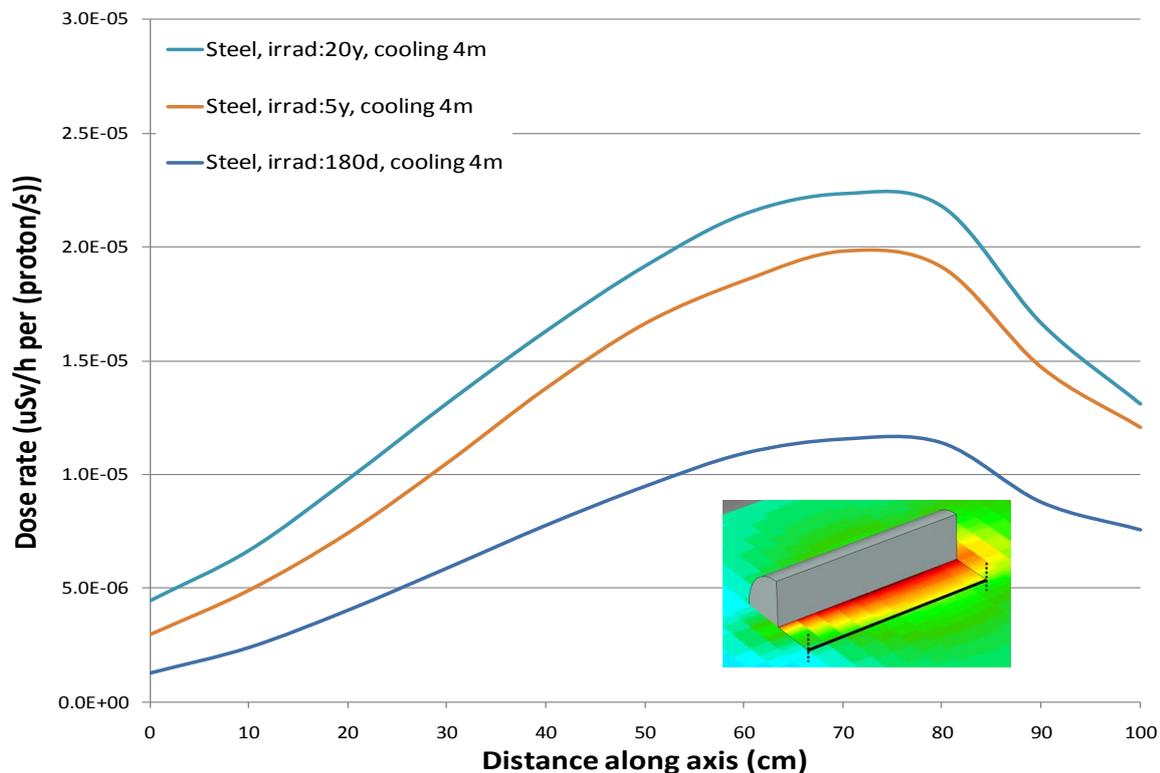

Figure 2: Results of calculations of ambient dose equivalent rates along an irradiated steel cylinder for irradiation times of 180 days, 5 years and 20 years.

The removal of the Inner triplet will imply destructive work on material with a relatively high level of radioactivity. Experience will be gained from the first triplet exchange in a few years from now. As mentioned above, the dose rates outside the magnets at Point 1 and Point 5 will be about 600 μSv/h after 4-month cooling and may reach much higher radiation levels inside the magnets. These values require a major optimization of the new generation of inner triplets with respect to design, installation, removal and transport. Material choice, flange connections and handling means need to be optimized.

The removal of collimators and warm magnets will lead to risk of workers' exposure to ambient dose equivalent rates in the order of some few 100 μSv/h up to mSv/h – even after four months of cooling. The dismantling of collimators was thoroughly studied and optimized and the development of a remote handling tool is well in progress. The dismantling of warm magnets and passive absorbers needs to be prepared and optimized – which is a priority for the next years of LHC operation. The installation of additional equipment to the already existing, radioactive material in Point 3 and Point 7 seems extremely difficult.

## RADIOACTIVE WASTE

The production of radioactive waste after 10 years of nominal operation of LHC was already estimated some years ago in the framework of the LHC waste study. After 20 years of operation, the waste production might turn out to be higher then the estimated one, because of increased intensities and luminosities and/or due to changes in the European legislation. CERN's present interim storage for radioactive waste, in the ISR tunnel, is not adapted to store LHC dipoles, also because of the lack of adequate means of transport. Therefore, a "light" storage solution for dipoles need to be studied. The dipoles might fulfill the acceptance criteria for low level waste in France and thus be eliminated towards the final repository Centre de stockage des déchets de très faible activité (CSTFA) in Aube. Radioactive waste others than dipoles will be temporarily stored at CERN - in shielded areas equipped with proper handling means - until elimination pathways are determined. It has to be taken into account that waste disposal regulation and techniques are likely to evolve over the next 20 years.

## CONCLUSIONS

Experience in removing components (dipoles, triplet, collimators) will already be gained in the next few years. The design of new components like the next generation of inner triplets needs to be optimized before being installed.

The radioactive waste production, storage and disposal should be addressed today – as even small amounts of radioactive waste from LHC risk to pose problems in view of handling, storage and elimination. The upgrade of the LHC to HE-LHC will increase the amount of radioactive waste. Options of recycling of components and material should be assessed, with a view to reduce the production of radioactive waste.